\documentclass[pdflatex,sn-chicago]{sn-jnl}

\usepackage{graphicx}%
\usepackage{multirow}%
\usepackage{amsmath,amssymb,amsfonts}%
\usepackage{amsthm}%
\usepackage{mathrsfs}%
\usepackage[title]{appendix}%
\usepackage{xcolor}%
\usepackage{textcomp}%
\usepackage{manyfoot}%
\usepackage{booktabs}%
\usepackage{algorithm}%
\usepackage{algorithmicx}%
\usepackage{algpseudocode}%
\usepackage{listings}%
\usepackage{comment}

\theoremstyle{thmstyleone}%
\theoremstyle{thmstyletwo}%

\theoremstyle{thmstylethree}%

\begin{document}

\title[Modeling revolutions in networked societies]{Modeling revolutions in networked societies: learning from the Tunisian spring}


\author[1]{\fnm{} \sur{Daniel Aguilar-Vel\'azquez}}\email{danielaguilar@fisica.unam.mx}
\author*[1]{\fnm{} \sur{Denis Boyer}}\email{boyer@fisica.unam.mx}
\author[2]{\fnm{} \sur{Robert Boyer}}\email{r.boyer2@orange.fr}


\affil[1]{\orgdiv{Instituto de F\'\i sica}, \orgname{Universidad Nacional Autónoma de M\'exico}, \orgaddress{\street{} \city{Mexico City}, \postcode{04510}, \state{} \country{Mexico}}}

\affil[2]{\orgdiv{\'Ecole des Hautes \'Etudes en Sciences Sociales (EHESS)}, \orgname{} \orgaddress{\street{54 boulevard Raspail}, \city{Paris}, \postcode{75006}, \state{} \country{France}}}



\abstract{Economic competition and deregulation have led to a polarization of societies between a small, increasingly powerful elite and a majority of socially excluded individuals, marginalized and unconnected to political representations. This is the breeding ground for protest movements, relayed by local ties and amplified by social networks. Based on the characteristics revealed by socio-economic research into the Arab revolutions of the 2010s, this article proposes a formalization inspired by the theory of complex systems. We discuss the conditions under which an initial localized event - for example, the suicide of a street vendor condemned to ruin in a small Tunisian town - can trigger an explosion of the number of opponents to the regime, typical of a revolutionary episode. We consider a network model of agents and oppressors where pair interactions are controlled by a fear parameter, or the inclination of the individuals to rebel despite of repression. The model exhibits a phase transition at a critical threshold above which the quiescent state becomes unstable. Furthermore, the ability of individuals to forge triadic relationships accelerates the process of joining a rising rebellion, making the transition more abrupt, in the form of a brutal discontinuity that can be described as revolutionary. The imposition of a counter-revolution by the hardening of repression can be explained by the hysteresis property displayed by the model: when mobilization has extended the initial network and made it partially permanent, repression by the authorities must be tightened to higher levels than before to regain control.
}




\maketitle

\section{Introduction}\label{sec1}

Understanding revolutionary processes has long been an ambition for history
\citep{hill1993revolution} and, by extension, for the social sciences. Revolutions can be social,
political, economic or technological, in configurations that vary in time and space.
The French Revolution never ceased to interest historians \citep{labrousse1949comment}, right
up to the present day. It could be explained by the emerging tensions within a very
particular society: feudal, agricultural and threatened by recurring food crises and
power struggles within the elite. It is undoubtedly this episode that forged the
definition of revolution as "an abrupt and violent change in the political structure
of the state that occurs when a group revolting against the existing authorities takes
power and succeeds in keeping it" \citep{larousse1986dictionnaire}. The Russian Revolution
obeyed this definition, but the actors were quite different, as society was organized
differently, which suggests that a general theory of revolutions is a long way off.
On the other hand, it is enlightening to look for the lineaments of a theoretical
model of the revolutionary phenomenon, based on the hypothesis of interactions
between horizontal processes linking actors and the verticality of power \citep{janne1960modele,stone2017theories}. However, even if this is an overstatement, in terms of
formalization, the models that were developed to account for the
economic dynamics of countries after the Second World War were essentially evolutionary \citep{nelson1985evolutionary,dosi2023foundations}, while revolutionary projects seemed to be abandoned. As a result,
economists involved in this evolutionary research program favored the analysis of
balanced growth paths.

The Arab revolutions that began in December 2010 have rekindled the interest
of social science researchers. Each discipline has contributed its own understanding
of certain aspects of these revolutions, and their conjunction can serve as a starting
point for an attempt at modeling, which is the purpose of the present article. 

Given the rapid spread and vast extent of Arab revolutions, that were largely unexpected, we re-examine here these events under the scope of emergence and cooperative phenomena. This approach, inspired from the physics of phase transitions and critical phenomena \citep{goldenfeld1992lectures,yeomans1992statistical}, generally aims to describe how the steady state of a large system composed of many interacting elements can suddenly loose its stability as a consequence of very small parameter changes or dynamical perturbations. In many systems, abrupt transitions toward collective behaviors embracing a large fraction of individual entities can occur spontaneously from apparently uncoordinated, local interactions among pairs, without the need of introducing explicit external factors or \lq\lq leaders". In these collective phenomena, a key role is also played by the network structure on which interactions take place \citep{newman2003structure,barrat2008dynamical}. This approach has proven to be very useful for understanding transitions to collective states in diverse contexts, such as epidemic outbreaks \citep{pastor2015epidemic}, the synchronization of coupled oscillators \citep{kuramosto1975self,strogatz1991stability}  or the self-sustained activity of neural networks in the brain \citep{kinouchi2006optimal}, to name a few. In the social sciences, equilibrium analysis shows that phase transitions may occur in principle in a wide range of systems, being particularly relevant for the understanding market crashes \citep{levy2005social}. 
The dynamics of phase transitions have also been discussed in the context of technological revolutions  \citep{malkov2023modeling} or evolutionary games of human cooperation \citep{perc2016phase}.

Another important aspect of the modeling approach adopted here is the assumption that interaction networks are not static but evolve in time, depending on the activity of the individuals that compose them. 
In changing social networks, old connections can fade or disappear with time, while new connections can be created between agents with similar traits and facilitated locally by common acquaintances, for instance  \citep{murase2015modeling, murase2019structural}. The 
co-evolution between individual activity and their social connections can lead to new behaviors, such as the enhancement of cooperative phenomena and stronger destabilization effects, as observed in opinion dynamics models \citep{holme2006nonequilibrium,iniguez2009opinion}, the formation of social networks \citep{marsili2004rise} or epidemic propagation \citep{hebert2015complex}.

This paper is organized as followed. In section \ref{sec:arabrev}, we recall the social context of Arab revolutions and pose three key questions of interest here. We then review previous modeling approaches that were inspired by the case of Tunisia and set the basics of a network model. In section \ref{sec11} we present this model and the stochastic rules that govern the individual dynamics. The results of numerical simulations are presented in section \ref{sec:results}. Section \ref{sec:disc} presents an interpretation of contemporary revolutions and we conclude in section \ref{sec13}.

\section{Arab revolutions, causes and specificities}\label{sec:arabrev}

A word of caution is in order, as observers originally used the term "spring"
rather than "revolution". In retrospect, after some successes, hopes of radical
transformation were dashed, to the point where other chroniclers later put forward
the idea of a winter of democracy. Another term, rebellion, could also be used, as it
does not prejudge the success of popular uprisings. Whichever term is chosen, the
same characterization applies: an episode marked by a sudden shift from isolated,
serialized individuals to a coalescence of a large fraction of the population against
the authoritarian elite, perceived as corrupt and illegitimate.

Most of the societies affected by the people's rebellion share a number of
common traits. The Arab revolutions are part of the end of post-colonialism
\citep{dabashi2012arab}, and are rooted in a demography marked by a large, young
population struggling to find employment \citep{gelvin2015arab}. The rise of social
networks, though uneven from country to country, multiplies the intensity of
connections between individuals \citep{lecomte2011tunisian}. These are authoritarian societies
in which the army is often called upon to arbitrate social and political conflicts
\citep{joshi2011arab}. The concentration of income and wealth pits a small elite against a
majority of the population. Poor economic prospects encourage the most dynamic
young people to emigrate, and barriers to immigration compound their frustration
\citep{oualdi2014ondes}. This proximity of social, political and economic configurations
undoubtedly explains the international transmission of the Arab revolutions and its
rapidity: Tunisia December 18, 2010, Algeria December 29, 2010, Jordan January
14, 2011, Oman January 17, 2011, Saudi Arabia January 21, 2011, Egypt January 25,
2011, Syria January 26, 2011, Yemen January 27, 2011, etc.

What are direct or indirect origins of these popular revolts? These are societies
shaped by the expectations of large numbers of young people who do not find their
place in economies whose development is mediocre in terms of living standards
and opportunities for expression in the political arena, due to the permanence in
power of authoritarian, elderly leaders who have appropriated a growing share of
national wealth \citep{filiu2011revolution,martinez2020state}.

Why did the Arab revolts occur at the end of 2010? In fact, the great American
financial crisis that began in 2008 has spread to all other countries, including
Tunisia. The slowdown in world trade further weakened economic prospects, while
soaring agricultural and food prices threatened the standard of living of the most
precarious: survival was at stake for a significant proportion of the population. The
Tunisian revolution was actually triggered by the action of one person: Mohamed Bouazizi, a 26-year-old itinerant
vegetable seller, immolated himself on December 17, 2010 in front of the governor headquarters of the small town of Sidi Bouzid. The anger spread to Kasserine, then to towns in
the Atlas Mountains and southern Tunisia, and finally reached the capital.

\subsection*{Tunisia: which models?}

The exemplary and precursory nature of the Tunisian revolution and the
availability of a study by Bouallegue that presents both the various economic models of
revolutions and an empirical analysis of their causes \citep{bouallegue2017analyse} inform our analyses. The author identifies the causes of this rebellion by
comparison with the 1980 hunger revolt. A principal component analysis (PCA)
reveals three major transformations: a demographic revolution characterized by the
importance of younger generations, the improvement of their level of education
and the raising of their expectations in the context of a very mediocre growth rate.

The key question, then, is how a "small", one-off, localized event can have
such far-reaching consequences, even in the long term. Indeed, in the majority of
economists' models, only a large shock can have a macroeconomic impact, albeit
only a transitory one, since stochastic general equilibrium models postulate the
structural stability of a long-term equilibrium \citep{lucas1972expectations}. More specifically, the
model in this article aims to account simultaneously for three stylized facts for which explanations are lacking:
\begin{itemize}
\item[1)] How, in the absence of a revolutionary organization or even a leader, the
interactions between purely individual decisions shifted political power? The
strengthening of trade unions and political organizations came after the
popular uprising.

\item[2)] Are certain types of interconnection and networks capable of generating a
brutal and cumulative transformation of the socio-political configuration?
What is the relationship between individual behavior, torn between the fear of repression and the urge to protest, and regime change?

\item[3)] Why, after a victory instituting more democracy, has repression finally
intensified compared to the pre-revolutionary period, with the return of an
authoritarian government?
\end{itemize}

As early as the 1970s, the "public choice" school proposed various models for
the revolutions observed up to that point. These models became rarer, until
the Arab revolutions revived interest among researchers. Which models are
most relevant to Tunisia?

$i)$ A first approach is to look for economic and social variables likely to
increase the probability of revolution. Tobit-type models thus mobilize
international data, assuming that determinisms are the same in all countries
\citep{mcdonald1980uses,marwell1993critical}. This method is not
available for the Tunisian case.

$ii)$ Another method is to estimate the determinants of the thresholds at which
an overall social movement is triggered \citep{kuran1989sparks,tullock1974social,tullock2005social}. However, without access to the threshold
distribution of individuals, we are forced to resort to a directly
macroeconomic approach, which does not allow to analyze the three stylized
facts under review, since individual interactions are not taken into account.

$iii)$ A central approach among economists is to apply the assumption of
rationality of behavior and of the formation of expectations about the
actions of other actors. The logic is that of collective action based on a
purely individualistic approach \citep{olsson2012evolutionary,palfrey1984participation}.
Those who believe they can derive economic benefits greater than the cost
of repression will engage in revolution. This presupposes calculating
capacities that exceed those of most participants in the rebellion movement.
In the Tunisian case, individuals are defending their standard of living, their
dignity, according to a limited rationality. Our modeling is following the
logic of evolutionary models, which are much more realistic and relevant.

$iv)$ A fourth approach is the application of game theory. In some cases, a two-
player strategic game is modeled, in which those in power oppose the
revolutionary activists. Or between rich elites and poor citizens \citep{acemoglu2000did}. In still other cases, three collectives clash: the ruling
elite, a protesting elite and the people \citep{roemer1985rationalizing}. Conceptually, these
models are interesting, but they are based on collective actors whose
emergence and organization have not been modeled. This does not apply to
the Tunisian case, which clearly shows the transition from isolated
individuals to a revolutionary crowd whose logic is no longer individualistic
and based on the pursuit of exclusively economic interests \citep{le1995psikhologiya}.
Then, as this author writes,  \textit{"a collective soul is formed, transitory no doubt, but with
very clear characteristics. The collectivity then becomes what, for want of a better
expression, " call an organized crowd, or, if you prefer, a psychological crowd. It forms a
single being, and is subject to the law of the mental unity of crowds."
}

In fact, a fifth avenue seems to be promising for our purposes. The transition
from individual to collective results from informational cascades when
individuals influence each other, without the intervention of a leader or
coordinator \citep{roemer1985rationalizing,banerjee1992simple,bikhchandani1992theory,borge2013cascading}. Can these horizontal interactions be
powerful enough to bring about a shift from a regime of isolated individuals
to a configuration where a collective is driven by a common goal? To answer
this question, the following formalization mobilizes the typical tools of the modeling of
complex systems.

\subsection*{The basics of the model}
An analysis of the causes of the Tunisian revolution has diagnosed a fertile
ground for the perception of a common interest for a large part of the population.
Firstly, to defend a standard of living that ensures survival; secondly, to protest
against the monopolization of income and wealth by a small elite holding political
power, including the power of repression; and thirdly, to democratize the political
regime. It is thus possible to characterize individuals by a parameter ($\sigma$ in section \ref{sec11} below) that measures
their capacity to share their discontent with other people, in the knowledge that they incur a probability of repression
that depends on the degree of coercion that governments can exert (related to the parameter $n$ in section \ref{sec11}).

The question then becomes: how do opinions form in contact with
neighboring individuals, either in strictly geographical terms, as the Tunisian case
shows, or through the power of social networks and modern forms of
communication? These observations justify the formalization of horizontal
interactions between individuals, which in fact have an unprecedented intensity in
connected societies. The proposed model suggests that the brutality of protest
movements stems from this characteristic. This is in stark contrast to the slowness
of transmission during the French Revolution, for example, due to the difficulty of
communications at the time.

In this respect, how do networks of geographical proximity and social
networks fit together? The proposed model shows that the spread of the revolt can be
qualitatively different, i.e., with a discontinuous jump of collective activity, when connected individuals not only
influence each other but can also establish new connections (see the probability parameter $p$ of section \ref{sec11}) to form triads and create small local communities. These connections can also disappear spontaneously with rate $m$, or through repression.
Incidentally, the chronology for Tunisia suggests a geographical spread, whereas
most analysts invoke the power of social networks. The fact that the spread of the
Internet has been extremely uneven from country to country puts this conclusion
into perspective \citep{cohen2011facebook,chenavaz2012printemps}.

The literature shows that the army played a decisive role in the outcome of Arab revolutions \citep{joshi2011arab}: very often the degree of repression was even greater
in order to defend the previous political order. This is a feature of the proposed
model: societies do not return to the status quo, but have become even more violent and repressive.


\begin{figure*}[h!]
\centering
\includegraphics[width=7.4cm,height=6.8cm, trim=.1cm .6cm .1cm .2cm, clip]{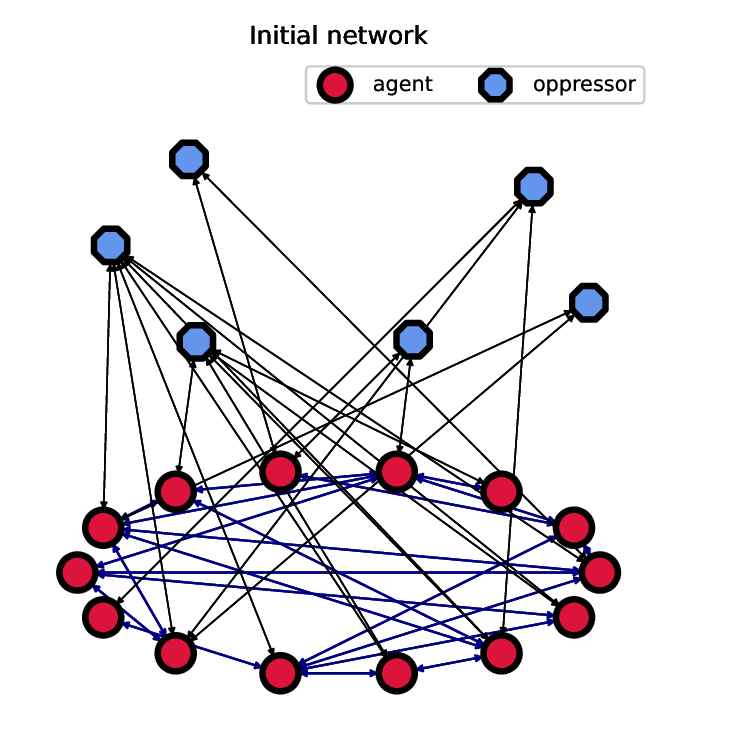}
\caption{Revolution network model. The network is composed of agents (lower layer) and oppressors (upper layer). Agents communicate with other agents and can be detected by oppressors. 
}
\label{fig1a}
\end{figure*}

\section{Methods}\label{sec11}

\subsection*{Network model}

The present model shares some similarities with a model by Kinouchi and Copelli developed in a very different context, namely, the perception capacity of sensory neural networks \citep{kinouchi2006optimal}. 
Our model considers a network of individuals which is composed of two layers. At the initial time, the first layer comprises $N$ nodes (hereafter called \lq\lq agents") forming a random 
network with average 
connectivity $K$ (Fig. \ref{fig1a}). To each agent $i$ is assigned $K_i$ connections to other agents of the same layer chosen at random, where $K_i$ is drawn from a Poisson distribution of mean $K$. These connections do not change over time. The second layer is composed of $N'$ nodes, hereafter called \lq\lq oppressors". Each oppressor node is connected 
to $K'$ randomly chosen agents of the first layer and these connections do not change over time either. 
In the following, we have set $N'=N/5$ and $K=K'$. 
Therefore, an agent is on average connected to $K/5$ oppressors. Other choices for the proportion of oppressors and $K'$ do not qualitatively change the results. 

\subsection*{Activation dynamics}

Each agent $i$ is described by a time-dependent, binary state variable $s_i(t)$ which informs about its rebellion activity. The inactive state is represented by $s_{i}=0$, while $s_{i}=1$ corresponds to the rebellion state, which is not necessarily disclosed to neighbors. All agents are discontented about their government and susceptible to enter in rebellion. Meanwhile, the oppressors in the second layer do not have a state {\em per se}, and their function is to always repress active agents. Time is discrete, or $t=l\Delta t$ with $\Delta t$ the time step and $l=0,1,2,\ldots$ an integer.

An agent can be activated (transit from $0$ to $1$) by essentially two mechanisms. The first one is spontaneous activation, modeled by a random Poisson process with rate $r$: at each time step $\Delta t$, an inactive agent $i$ can undergo the transition $0\to1$ with probability $\lambda=r\Delta t$. We will consider the situation where such spontaneous activation is very unlikely: we choose $\lambda=1/N$, therefore about one individual on average rebels spontaneously at a given time. The second mechanism corresponds to activation through neighbors:
an agent $j$ which is active at time $t-\Delta t$ and has a connection with an inactive $i$ activates the latter at time $t$ with probability $a_{ij}$. The probabilities $a_{ij}$ are fixed random variables with uniform distribution in the interval $[ 0 , a_{max} ]$, where $a_{max}$ is a constant parameter. The interaction matrix is symmetric, or $a_{ij}=a_{ji}$

The agents individually remain little time in the active state in order to escape oppression. Hence, an agent which is active at a given time (through either mechanism) returns to the inactive state $s_{i}=0$ in the next time step.

The activation strength $a_{max}$ is linked to the fear felt by the agents: when individuals are afraid of communicating their state to neighbors, $a_{max}\ll 1$, while $a_{max}$ close to unity mimics a situation where communication is open. We define the so-called local branching ratio, $\sigma_{j}=\sum_{i}^{k_{j}} a_{ij}$, which represents the
average number of activations that
the $j$th element can create in one time step, given that it is connected to $k_j$ other agents. 
The main parameter of the model is the average of $\sigma_j$ over all agents and is denoted as $\sigma$. This parameter can be used instead of $a_{max}$ through the relation $\sigma=Ka_{max}/2$. Although other independent parameters refer to the repression process (see below), in a population of unsatisfied individuals, the fear parameter $\sigma$ itself can be seen as a proxy of repression: at equal social/economical conditions, $\sigma$ is small when repression is severe and larger when agents feel safer expressing their opinion to others.


\subsection*{New links rule}
\begin{figure*}[h!]
\centering
\includegraphics[width=7.4cm,height=6.8cm, trim=.1cm .6cm .1cm .2cm, clip]{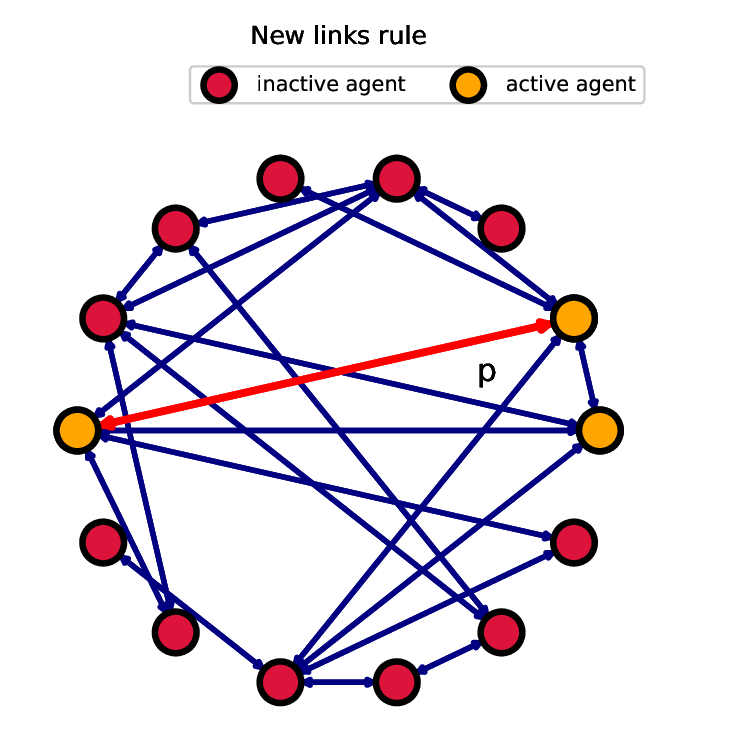}
\caption{New links rule. In the lower layer (shown alone for clarity), new links in red can be formed with probability $p$ between two agents, provided that both agents are in the active state and share a common neighbor that is also active. Any new connection is removed with probability $m$ at each time-step, and kept with the complementary probability $1-m$. 
}
\label{fig1b}
\end{figure*}

New connections can be created between pairs of initially unconnected agents when both of them are active and share a common neighbor that is also active (Fig. \ref{fig1b}). Under this condition, the new connection is created with probability $p$. This rule leads to the formation of triangles in the network or {\em triadic closures}.
If created, the new link between $i$ and $j$ is assigned a fixed random number $a_{ij}=a_{ji}$, chosen uniformly in the interval $[0,a_{max}=2\sigma/K]$, as for the activation probabilities in the initial network.

However, these new connections have a finite lifetime: at each time step, new connections are removed with probability $m$. In contrast, the $K$ initial connections that each individual had in the original network remain permanent. A maximum of 100 connections are allowed for each agent, which is of the order of the Dunbar number, the empirical maximal number of relations an individual can maintain in a society \citep{dunbar1992neocortex}.



\subsection*{Oppression mechanism}
Similarly to the activation processes among agents, an active agent $i$ can be detected with a fixed probability $a_{\ell i}$ by an oppressor $\ell$ to which it is connected. This probability is also uniformly distributed in $[0,a_{max}]$. If at least one oppressor becomes aware that $i$ is active during a time-step, the agent enters in a refractory state (e.g., put in jail). In addition, all the new links that $i$ had established through triadic closure are eliminated. The repressed agent stays in the refractory state during $n$ consecutive time steps and cannot activate other agents during that period. At the end of the refractory period, $i$ returns to the inactive state $s_i=0$ and resume the dynamics of the non-repressed agents with its $K$ original neighbors.


\subsection*{Simulations and main quantities}

In summary, the main parameters of the model are $\sigma$ (inversely correlated to fear), $p$ (new link creation), $m$ (new link removal) and $n$ (duration of the repressed state). We have performed Monte Carlo simulations of the model with a network of $N=10^{5}$ agents (hence $2 \times 10^{4}$ oppressors) with mean degree $K=10$. The spontaneous rebellion probability is $r=10^{-5}$. Unless indicated, at the initial time $t=0$, all agents are set in the inactive state $\sigma_i=0$, with their $K$ initial random connections, on average. 
The time $t$ is incremented of one unit ($\Delta t=1$ from now on) when $N$ individual states have been updated, one after the other (asynchronous updates). The total number of time-steps in a simulation is $10^3$.

To characterize the macroscopic dynamics of the system, we computed the time-dependent average
activity given by 
\begin{equation}\label{rho}
\rho({t})=\frac{1}{N}\sum_{i=1}^{N}{s_{i}(t)}. 
\end{equation}
We will mainly focus on the fraction $F\in[0, 1]$ of active individuals at late times, through the study of the time average of $\rho(t)$ defined as
\begin{equation}\label{F}
F= \frac{1}{T}\sum_{t=s}^{s+T} \rho({t}), 
\end{equation}
where $T$ is a measurement interval and $s$ a large enough time.

Basic network quantities are also of interest, such as the average degree $k(t)=\sum_{i}^{N} k_i(t)/N$ where $k_{i}(t)$ represents the number of neighbors of agent $i$ at time $t$. We have also monitored the clustering coefficient of the network, i.e., the probability that two nodes connected to a same node are also connected to each other (transitivity) \citep{newman2003structure}. The clustering coefficient for a agent $i$ is defined as,
\begin{equation*}
    C_{i}(t)=\frac{1}{k_{i}(t)[1-k_{i}(t)]} \sum_{j,k}A_{i,j}(t)A_{j,k}(t)A_{i,k}(t),
\end{equation*}
where the coefficient of the adjacency matrix $A_{i,j}(t)$ indicate whether there is a connection between agents $i$ and $j$: $A_{i,j}(t)=1$ if they are connected, otherwise $A_{i,j}(t)=0$. The average clustering coefficient $C(t)$ is obtained by taking the average of $C_i(t)$ over the $N$ agents.




\section{Results}\label{sec:results}
\subsection*{Phase transition in the steady state}

\begin{figure*}[h!]
\centering
\includegraphics[width=7.4cm,height=6.8cm]{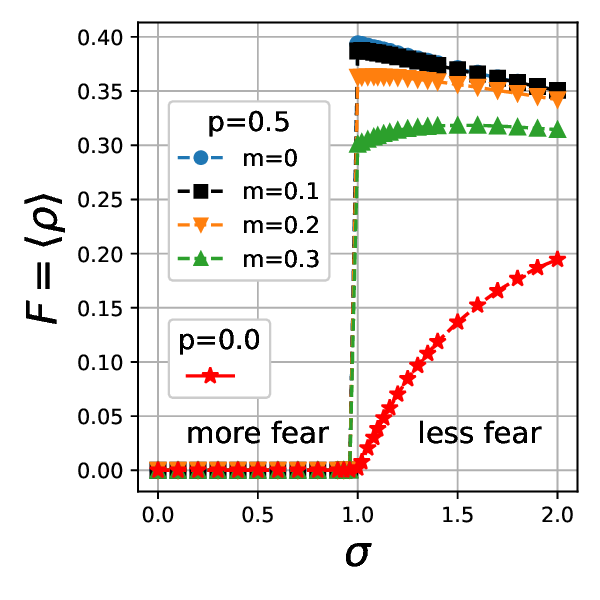}
\caption{Fraction of active individuals in the steady state as a function of the fear parameter $\sigma$ (other parameters are $N =10^{5}$,  $n=5$, and $K=10$), Each dot represents an independent run. 
}
\label{fig2}
\end{figure*}

In the first scenario studied, the new links are created easily among active agents through triadic closure and have a rather long lifetime, i.e., the probability parameter $p$ is large (set to $0.5$) whereas link removal is relatively small ($m=0;0.1$). 
Fig. \ref{fig2} displays the fraction of active individuals in the steady state at late times, $F$, as a function of the fear parameter $\sigma$ (recall that smaller values of $\sigma$ correspond to agents with little communication). 
Despite of the fact that the probability of spontaneous rebellion is very small ($r=10^{-5})$, the network is able to reach a macroscopic rebellion state with $F\sim 0.2-0.4\gg r$. An abrupt change takes place when $\sigma$ crosses a critical value $\sigma_c\simeq1$. When $\sigma<\sigma_c$, $F\simeq0$ and only a few individuals revolt sporadically (quiescent phase). If $\sigma>\sigma_c$, on the contrary a large fraction of the population sustains active rebellion (revolutionary phase). This high activity persists even if the link removal probability is relatively large ($m=0.3$).

Importantly, the steady state activity 
is {\em discontinuous} at the transition. This means that a slight change in $\sigma$ around $\sigma_c$ produces a jump of finite magnitude in the collective activity of the system. In the Ehrenfest classification scheme of phase transitions, this behavior would correspond to a \lq\lq first-order"  transition. Here, the discontinuity is essentially a consequence of the creation of new links in response to the activity of the agents. 

We contrast these results with a second scenario, where the creation of new links does not occur, hence the interaction network does not evolve with time and stays in its initial configuration ($p=0$ while $m$ is irrelevant). In this case, a sharp phase transition occurs at $\sigma_c\simeq1$, too, but it is smoother and continuous (\lq\lq second-order" phase transition in the Ehrenfest scheme), as shown in the red curve Fig. \ref{fig2}. In this case $F$ grows continuously from zero for $\sigma\ge\sigma_c$, although $\partial F/\partial \sigma$ undergoes a discontinuity at $\sigma=\sigma_c$. Continuous or second-order phase transitions are standard on static networks and occur in models of neuronal activity \citep{kinouchi2006optimal} or epidemic spread \citep{pastor2015epidemic}, for instance. Note that in both types of transition, the singular behaviors at $\sigma=\sigma_c$ are reached in principle in the limit $N\to\infty$ and $r\to0$. In fact, the steady states represented in Fig. \ref{fig3} do not depend on $r$, as long as it is very small.

\subsection*{Uprising dynamics at the transition point}

\begin{figure*}[h!]
\centering
\includegraphics[width=7.4cm,height=6.8cm]{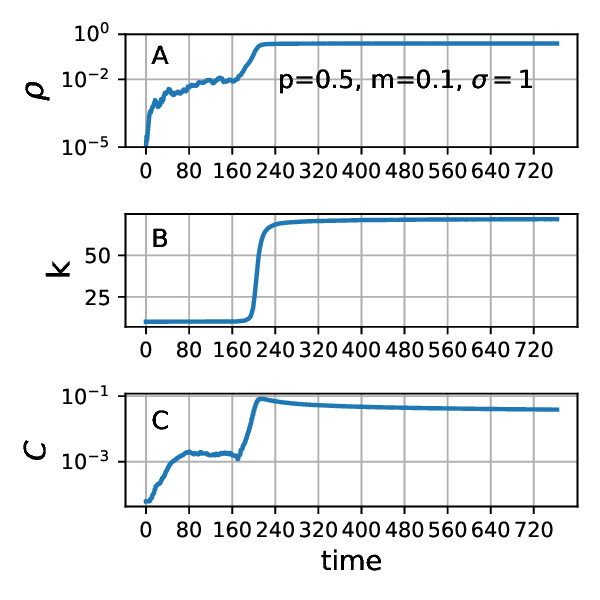}
\caption{Example of time series of activity $\rho(t)$, connectivity $k(t)$, and average clustering $C(t)$ in a single run at the transition point $\sigma=1$.
}
\label{fig3}
\end{figure*}

It is also instructive to monitor the evolution of the density $\rho(t)$ of active agents with time in a single numerical experiment performed at $\sigma=\sigma_c$ (Fig. \ref{fig3}-A). 

The system rapidly reaches a first plateau of low activity ($\rho(t)\sim 0.1-1\%$), while the mean degree of the network $k(t)$ remains very close to its base value $K$ (Fig. \ref{fig3}-B). The density of active agents is nevertheless much larger than $r$, hence rebellion occurs through interactions, but their frequency is not sufficient to produce many new connections, as it is unlikely that two active agents share a common neighbor, or because they are repressed. After a certain time, the uprising itself takes place, consisting of a roughly exponential growth of $\rho(t)$ paralleled by a rapid growth of $k(t)$. This phase is short (about $40$ time units) and followed by the steady state regime with $F\sim 40\%$. 

The evolution of the network can also assessed through the clustering coefficient $C(t)$, whose variations are qualitatively similar to that of $\rho(t)$ (Fig. \ref{fig3}-C): it is very small at $t=0$ (of order $1/N=10^{-5}$ due to the randomness of the initial network \citep{newman2003structure}) and rapidly grows to an intermediate value of $\sim 10^{-3}$, indicating that a few new connections are actually being created through triadic closure and able to sustain a small activity. $C(t)$ then rapidly grows and reaches a much larger value close to $0.1$. Hence, the structure of the network has changed significantly from its initial configuration.



\subsection*{Hysteresis effects and the intensification of 
oppression}

\begin{figure*}[h!]
\centering
\includegraphics[width=12.4cm,height=8.8cm]{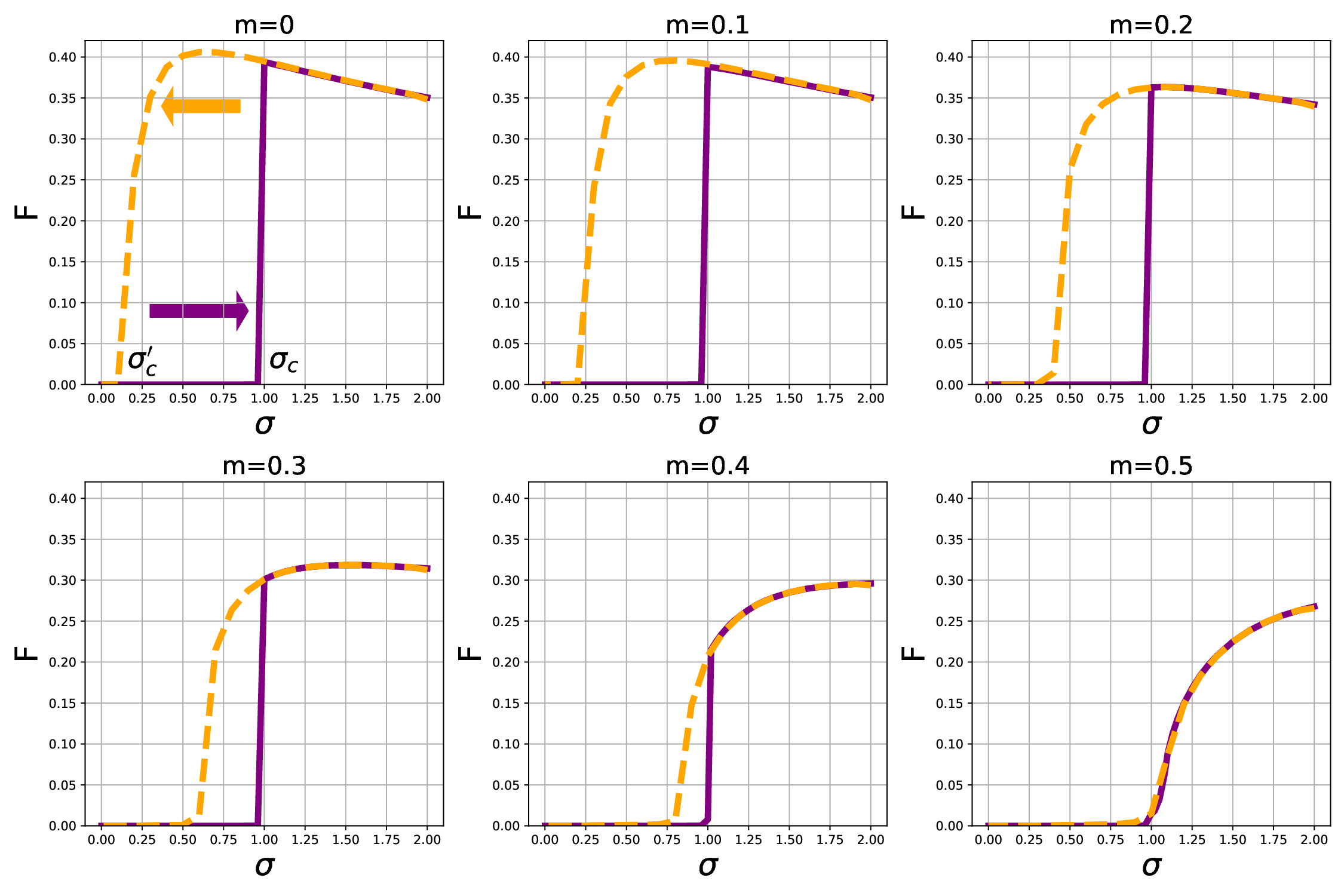}
\caption{Hysteresis curves of the  for $p=0.5$ and several link deletion probabilities $m$. The parameter $\sigma$ is gradually increased from $0$ to $2$ and back (see text).}
\label{figure2}
\end{figure*}

Discontinuous transitions are in general closely related to the phenomenon of hysteresis, or resistance to change, which leads to non-reversible responses in systems subject to external drives \citep{goldenfeld1992lectures}. 
Here, we relate hysteresis to the intensification of oppression after a massive uprising, beyond the oppression levels in place before the revolution [see point 3) of section \ref{sec:arabrev}]. Let us consider the following numerical experiment: we simulate a single network starting with $\sigma=0 <\sigma_c$ and increase $\sigma$ by small increments every $10^3$ time-steps of the dynamics, until reaching $\sigma=2$, an arbitrary value beyond the transition point $\sigma_c=1$. We do not reset the states or connections of the agents each time $\sigma$ is changed. The increase of $\sigma$ may be associated to worsening economical indices, a changing demography as discussed in section \ref{sec:arabrev}, or to a decrease of fear caused by a perceived weaker oppression. The behaviors of $F$ vs. $\sigma$ are represented as solid purple curves in Fig. \ref{figure2}, for several values of the link removal parameter $m$. These curves are practically identical to those of Fig. \ref{fig2}.

\begin{figure*}[h!]
\centering
\includegraphics[width=8.cm,height=5.3cm]{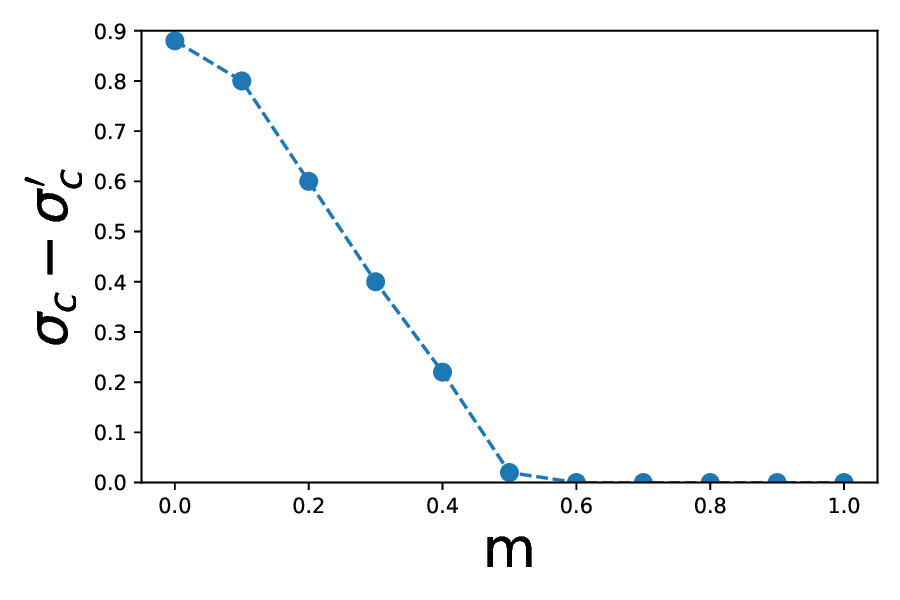}
\caption{Dynamical robustness $R=\sigma_{c}-\sigma_{c}'$ as a function of the deletion probability of the new links $m$ for the curves of Fig. \ref{figure2}.}
\label{figure3}
\end{figure*}

When $\sigma>\sigma_c$, a large fraction of agents are active, their network has been re-organized, and we can analyze under which circumstances the system can be forced to return to the quiescent state with $F \simeq 0$. To do so in a simple way, we proceed to bring $\sigma$ back to $0$ by small decrements every $10^3$ time steps, starting from $\sigma=2$ (orange dashed curves of Fig. \ref{figure2}). This protocol can model an increase of fear, seen as a proxy of intensified oppression after massive rebellion. 
As shown for instance by the case $m=0$ of Fig. \ref{figure2}, the system exhibits hysteresis, i.e., the orange curve of the path $\{\sigma=2\}\to\{\sigma=0\}$ does not match the previous curve $\{\sigma=0\}\to\{\sigma=2\}$. This feature first confirms the discontinuous nature of the transition seen in Fig. \ref{fig2}, as continuous transitions do not exhibit hysteresis and are reversible \citep{goldenfeld1992lectures}. Secondly, the agents resist oppression much better on the way back, where $F$ can remain high even for $\sigma<\sigma_c$, a region where it was vanishing.
It thus takes more efforts (higher fear levels) to return to the quiescent state $F\simeq 0$. 


This effect is particularly strong when the new connections formed by triadic closure are long-lived, as in the cases $m=0$ and $m=0.1$. In these examples, it is possible to sustain a substantial rebellion activity even at very small $\sigma$ (Fig. \ref{figure2}, e.g., case $m=0$). Remarkably, in those cases, the global activity $F$ can be larger when the interaction between agents {\em weakens} (or fear increases). For instance, as $\sigma$ decreases, the orange curve of $F$ reaches a maximum at some value which is smaller than $\sigma_c$. Therefore, the highest levels of activity are seen in the vicinity of $\sigma_c$ and not for larger values, as one would naively expect and as observed for other cases with larger $m$.

When the new connections become short-lived ($m$ increases), the discontinuous jump at $\sigma_c$ is smaller and  hysteresis is less pronounced. At a particular value $m=0.5$, the discontinuity disappears altogether and the dotted and solid lines are superimposed, indicating the absence of hysteresis. In this case the phase transition is continuous, similarly to the case $p=0$ (no new connections created).



Hysteresis is directly related to the dynamical resilience 
of the network and can be quantified through $\sigma_{c}'$,  
the value of $\sigma$ at which 99$\%$ of the population has returned to the inactive state in the path $\{\sigma=2\}\to\{\sigma=0\}$. 
As clear from Fig. \ref{figure2}, when the new connections live shorter ($m$ increases), 
$\sigma_c'$ approaches $\sigma_c$. 
We define the hysteresis strength or dynamical resilience as $R=\sigma_{c}-\sigma_{c}'\ge0$. 
The variations of $R$ with the link removal parameter $m$ are shown in Fig. \ref{figure3}. In this example, if $0\le m<0.5$, the transition is discontinuous and $R>0$, whereas if $0.5<m\le 1$, the transition becomes continuous and $R\simeq0$.


\subsection*{Varying the oppression parameters}

Transition to collective rebellion occurs when the branching ratio overcomes the value $\sigma_c\simeq 1$, i.e., when an active agent influences on average one or more individuals in one time-step. By performing simulations with many parameter choices, one observes that $\sigma_c$ is actually always unity and thus independent of the parameters of the model such as the refractory time $n$. This transition thus share some similarities with the critical dynamics of branching processes \citep{marro1999noequilibrium}. However, the oppression parameter $n$ does have an impact on the values of $F$ reached {\em after} the transition ($\sigma>\sigma_c$), when many agents are actually being repressed. 

\begin{figure*}[h!]
\centering
\includegraphics[width=12.4cm,height=6.8cm]{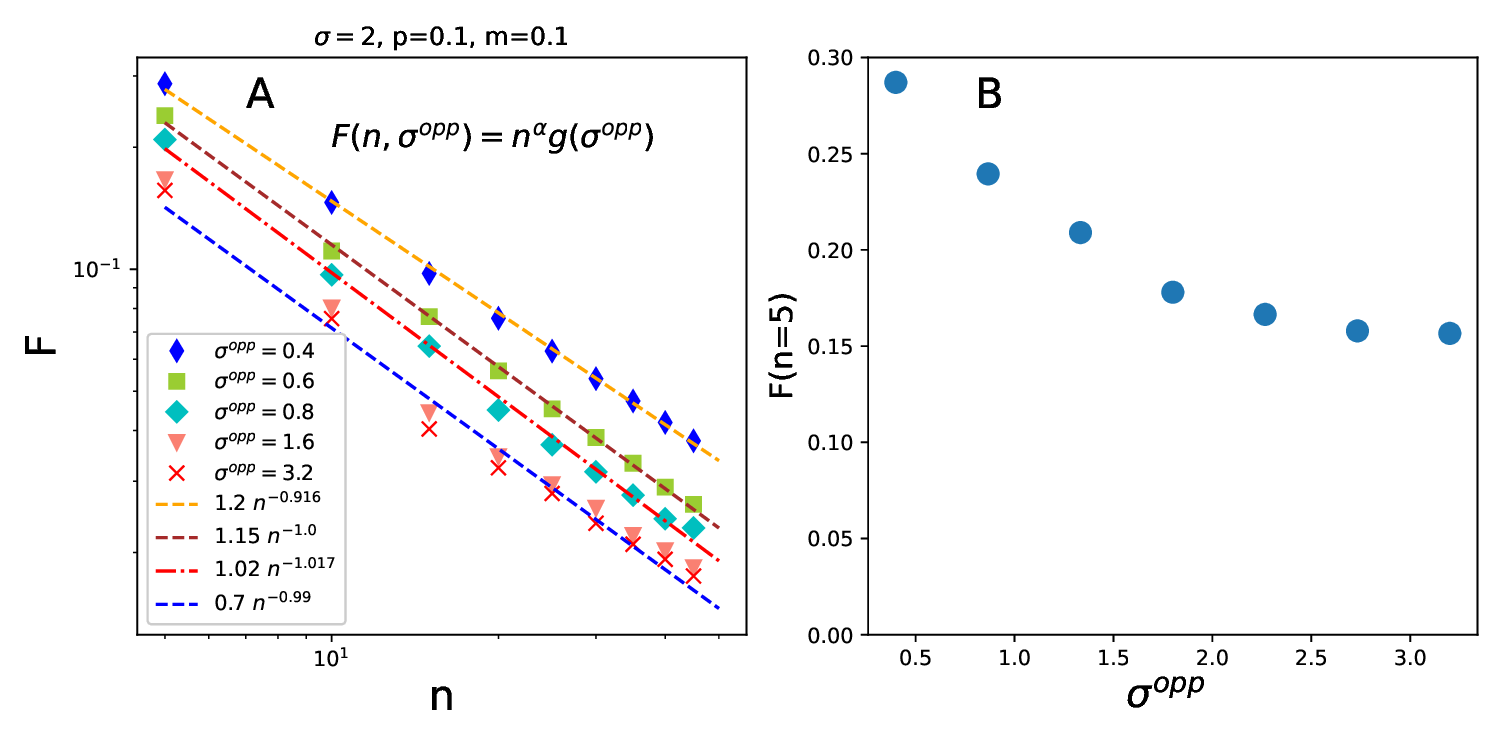}
\caption{Collective activity $F$ vs. the refractory period $n$ (A) and detection parameter $\sigma^{opp}$ of active agents by oppressors (B). We set $\sigma=2.0$ among agents, $p=0.1$, and $m=0.1$. When active agents are repressed for longer times $n$, the system gets close to the quiescent state $F\simeq 0$. However, easier detection by oppressors (higher $\sigma^{opp}$) results in a mild decrease of $F$.  
}
\label{figure4}
\end{figure*}

Fig. \ref{figure4}A shows how the activity $F$ at $\sigma=2$ decreases with $n$ and tends to $0$ at large $n$, as $F\sim 1/n$. Hence, bringing the system closer to the quiescent state after the revolution (while keeping $\sigma$ constant, unlike in the hysteresis experiment) requires an increased duration of imprisonment. A finite $n$ does not completely eliminate rebellion, though.

A third way by which $F$ can be affected in the steady state is through a modification of the detection capacity of active agents by oppressors. Let us introduce a parameter $\sigma^{opp}\neq \sigma$ which is specific to the connections agents/oppressors. Hence, instead of choosing the probability of repression of agent $i$ by $\ell$ as a uniform random number in the interval $[0,2\sigma/K]$, it is chosen in the interval $[0,2\sigma^{opp}/K]$. If $\sigma^{opp}>\sigma$, oppression actually becomes more efficient than the communication between agents. Fixing $\sigma=2$ and $n=5$, the variations of $F$ with $\sigma^{opp}$ exhibit a decay followed by a saturation at large $\sigma^{opp}$. Hence a more efficient detection alone is not sufficient to bring collective activity toward very low levels.



\subsection*{Robustness of the results}

The rules of the model can be changed in many ways, still, the main properties discussed above are quite robust. The figures \ref{fig2}-\ref{figure4} are not qualitatively affected by the following modifications or extensions:
\begin{itemize}

\item [] (a) the agents stay active during more than one time step before returning to $s=0$;

\item [] (b) new links are not removed when an agent is put in the refractory state;

\item [] (c) the interaction matrix is not symmetric, or $a_{ij}\neq a_{ji}$;

\item [] (d) the creation of new links is not restricted to triadic closure: at each time step and for each active agent, one new connection is established (with probability 1) with another active agent chosen randomly in the network, i.e., not necessarily sharing a common neighbor. 

\end{itemize}

\begin{figure*}[h!]
\centering
\includegraphics[width=12.4cm,height=6.8cm]{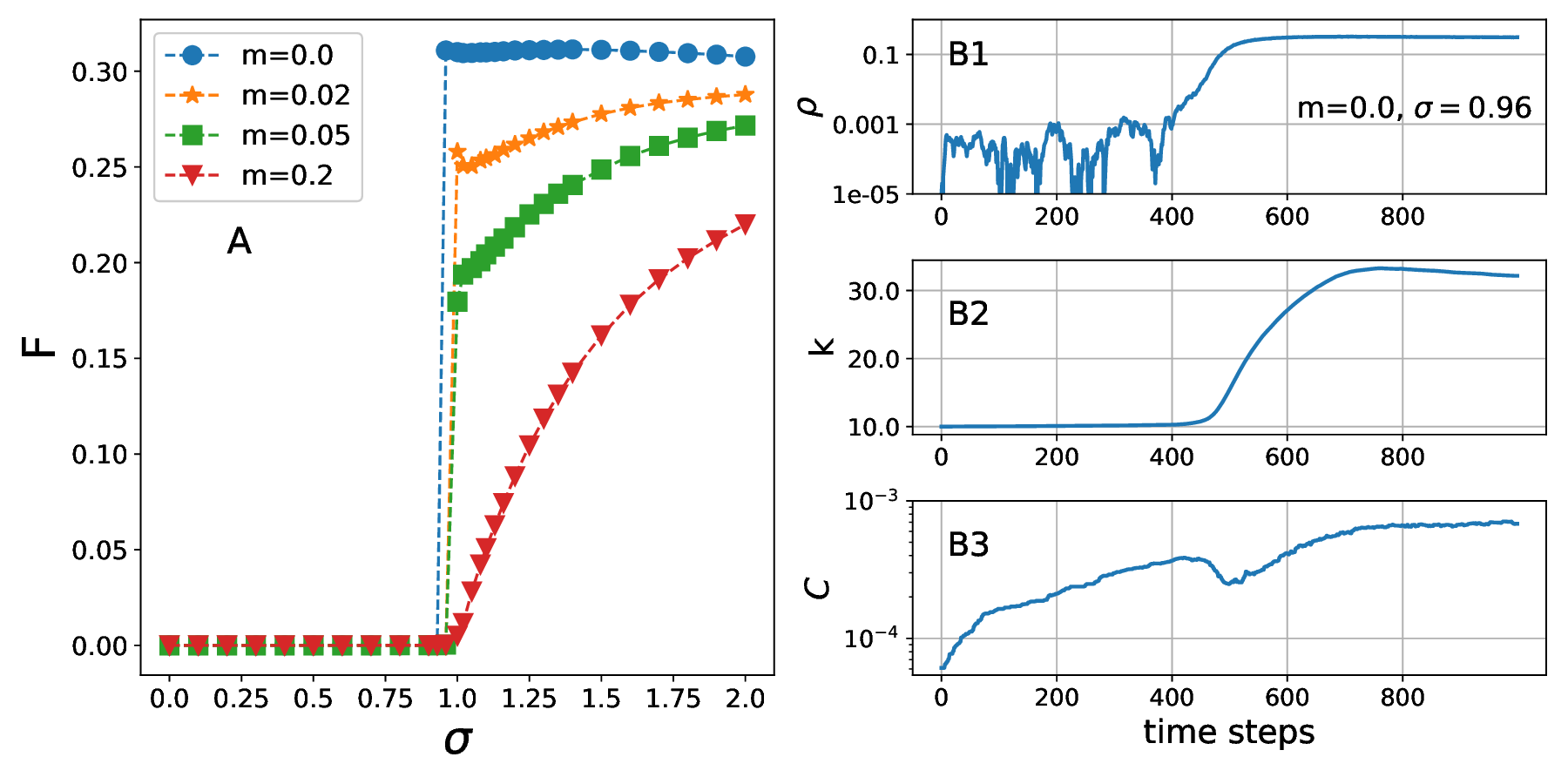}
\caption{New connections are created randomly between active agents, instead of by triadic closure. A) Steady state activity for different link removal probabilities $m$. Time evolution of the activity (B1), mean degree (B2) and clustering (B3) at $\sigma=0.96$ and $m=0.0$. A massive activity can be reached at a lower connectivity ($k\sim 30$) than in the triadic closure rule but the average clustering coefficient remains $<10^{-3}$. }
\label{figure5}
\end{figure*}

For instance, the variant (d) might be relevant to modern uprisings in which discontented people form bounds through the Internet without belonging to a same neighborhood or community \citep{howard2011opening}. In Fig. \ref{figure5}A we show the collective activity of the system in that case for different values of $\sigma$ and $m$. As in the previous version, we observe a discontinuous phase transition for $m=0$ and $m=0.05$, but the transition is continuous for $m=0.2$. The phase transition also occurs at $\sigma_c$ close to $1.0$. The time evolution of $\rho(t)$ also follows the three main regimes identified in Fig. \ref{fig3}A: pre-revolutionary low activity, exponential growth and steady state regime  (Fig. \ref{figure5}B-1). However, the connectivity $k\sim 30$ reached in the steady state is significantly lower (Fig. \ref{figure5}B-2). Due to the suppression of the triadic closure condition, the average clustering coefficient $C(t)$ is two decades lower than in the previous version (Fig. \ref{figure5}B-3).

\section{An interpretation of contemporary revolutions}\label{sec:disc}

Our modeling highlights the role of four key parameters: the individual propensity to revolt (i.e., the inverse of the level of fear of repression), the ease of creating new triadic links, the speed of losing new links, and the duration of imprisonment of individuals affected by repression. The combination of these factors enables us to characterize some of the processes that took place in Tunisia and other revolutions.

$i)$ Ex ante coordination of individuals is not necessary for a society to tip over into rebellion against the authorities. The general context, both economic (threat to the survival of the most disadvantaged) and political (authoritarianism, corruption, denial of individual rights), can create a threshold such that the rebellion of a single individual triggers a cumulative shift in attitudes.

$ii)$ As the fear parameter varies, this movement is characterized by a discontinuous jump that involves a large fraction of the agents, provided that new links can be created and are sufficiently persistent over time. Conversely, if the activation dynamics take place on a static network or if new links are shot-lived, the crossing of the same uprising threshold results in a gradual increase in the number of rebels, without discontinuity. It would seem that the Tunisian trajectory belongs to the first configuration, since, for example, trade unions, political parties and social movements took over and in turn developed new connections.

$iii)$ Leaders, driven by their own interests, are not building collectives to organize actors for the revolution in the early stage. Rather, individuals embedded in local links trigger an avalanche of
information that restores the role of collective organization. In both cases, there is no rational calculation of the benefits and costs of joining the revolutionary process, which is at odds with microeconomic logic.

$iv)$ These same mechanisms also explain the third stylized fact at the origin of this research. Once the co-evolution between players' activity and the dynamics of new links have created a stable steady state, it is not enough to exercise the same coercion as in the past to return to a quiescent society. 
Repression has to be even stronger, and this is the common feature of all Arab rebellions: none has succeeded, and even more authoritarian governments have emerged from them. The army is often the arbiter that imposes a return to order, like in Egypt.

$v)$ The hypothesis that online social networks are the source of the transformation of social and political movements that have become explosive due to the speed of modern means of communication can also be put into perspective. On the one hand, our modeling also qualitatively confirms the existence of a threshold at which a steady state of rebellion emerges abruptly when agents connect randomly. On the other hand this mechanism is not a necessary condition, as triadic links formed from geographical proximity allow at least as large uprisings. This once again highlights the specificity of the Tunisian case.

This model is just a first step towards understanding certain contemporary revolutionary episodes.

\section{Conclusions}\label{sec13}

The literature agrees in defining revolutions as a brutal and violent change in the political order when a group revolts against the authorities in order to seize power. As such, the Arab Spring, which began in Tunisia, falls into this category. Yet, they are in no way a repetition of emblematic events such as the French revolution of 1789 or the Russian revolution of 1917. Indeed, the latter were the result of parties and organized social groups fighting against a state that defended the interests of an elite. In contrast, it seems that a localized event affecting a single individual has triggered the tipping of the Tunisian society against the authorities. Only afterwards organized collectives (parties, unions and associations) defended the revolution according to their interests, surfing on the mobilization of citizens.

This article proposes a model inspired by the dynamics of complex systems, whereby interconnected agents can give rise to the emergence of collective phenomena. Models of this type are useful as they propose microscopic mechanisms for the destabilization of a macroscopic steady state beyond a certain threshold, causing that a single event is enough to cause a society to abruptly shift from one configuration to another. Abrupt transitions have been found in models of opinion dynamics with bounded confidence \citep{deffuant2000mixing}, or in the structure of networks whose evolution is driven by triadic closure \citep{marsili2004rise}. The Axelrod's model for the formation of cultural domains can also display a discontinuous transition in parameter space when interactions are short-ranged \citep{castellano2000nonequilibrium}. Discontinuous transitions are also observed in social contagion processes on hypergraphs \citep{skardal2020higher,de2020social}, in which a hyper-edge represents a grouping of two or more agents, and where the number of active agents drives a threshold process \citep{de2020social}. In neurosciences, similar phase transitions have been related to hyper-excitable systems, like epilepsy \citep{wang2023critical}. First-order transitions have been reported in neural network dynamics, and are promoted principally by an increase in the strength of connections \citep{scarpetta2018hysteresis,bachtis2020mapping}.
A configuration of fully active agents can also be achieved in networks
in which the most connected nodes are also connected to each other, forming a rich club \citep{aguilar2019critical,aguilar2021critical}.

Our model makes explicit some of the conditions that are sufficient to ensure an abrupt shift in rebellion. Firstly, an economic and political context that marginalizes a fraction of the population, ready to protest even in the face of repression by the authorities. Secondly, the ability to forge triadic relationships at local level, which accelerates the process of joining an open uprising. There is indeed the possibility of a brutal discontinuity of the kind observed in various Arab societies. Our results thus highlight the key role played by network co-evolution in parallel to communication between pairs of individuals: the densification of the interaction network not only allows a discontinuous transition but also confers resilience to the system in front of increased oppression.
The latter property of the model is consistent with observations: when
mobilization has exceeded the initial network and partially perpetuated it, repression must be tightened to regain control by the authorities. In fact, even more authoritarian governments were imposed after the Arab revolutions.

This approach should be extended in several directions. To what extent do networks of geographical proximity and virtual social networks affect the capacity for political mobilization differently, a theme that has only been partially addressed? The social networks seem to have encouraged the emergence of electoral coalitions with no coherent political or economic program, rather than the revolutionary initiatives by citizens. In the context of economics research, social networks seem to have been decisive in some revolutions, for instance concerning the shift from Keynesian to neoclassical conceptions of economic policy, given the density of the  network formed by economists at an international scale.


These considerations are in line with the relationship between history and modeling. On the one hand, theorists often focus on the mechanisms that ensure an economy structural stability. On the other hand,
modelers are called upon to explore both the emergence of new configurations and major crises. In both cases, the aim is to understand regime shifts that evolves over much shorter time scales from those which prevail within a stabilized regime. Such episodes are less frequent, but they are crucial and their intelligibility is a challenge. On the other hand, revolutions and crises have forms that are indexed to the socio-economic regime, the political system and the geography of networks. According to this vision, it is probably illusory to seek a generic formalization valid in all times and all places.

\bmhead{Funding}

D.A.V. thanks Secretaria de Ciencia, Humanidades, Tecnolog\'\i a e Innovaci\'on (SECIHTI) for their support with the postdoctoral fellowship \lq\lq Estancias Posdoctorales por M\'exico 2023".

\section*{Declarations}

\subsection*{Code availability}
The code in C++ language to produce the results is available at github:
https://github.com/danielvelaguil/Revolution-network-model


\begin{thebibliography}{}
\providecommand{\doi}[1]{\url{https://doi.org/#1}}
\bibcommenthead

\bibitem[\protect\citeauthoryear{Acemoglu and Robinson}{Acemoglu and Robinson}{2000}]{acemoglu2000did}
Acemoglu, D. and J.A. Robinson. 2000.
\newblock Why did the west extend the franchise? democracy, inequality, and growth in historical perspective.
\newblock {\em The quarterly journal of economics\/}~{\em 115\/}(4): 1167--1199 .

\bibitem[\protect\citeauthoryear{Aguilar-Vel{\'a}zquez}{Aguilar-Vel{\'a}zquez}{2021}]{aguilar2021critical}
Aguilar-Vel{\'a}zquez, D. 2021.
\newblock Critical neural networks minimize metabolic cost.
\newblock {\em Physics\/}~{\em 3\/}(1): 42--58 .

\bibitem[\protect\citeauthoryear{Aguilar-Vel{\'a}zquez and Guzm{\'a}n-Vargas}{Aguilar-Vel{\'a}zquez and Guzm{\'a}n-Vargas}{2019}]{aguilar2019critical}
Aguilar-Vel{\'a}zquez, D. and L.~Guzm{\'a}n-Vargas. 2019.
\newblock Critical synchronization and 1/f noise in inhibitory/excitatory rich-club neural networks.
\newblock {\em Scientific reports\/}~{\em 9\/}(1): 1258 .

\bibitem[\protect\citeauthoryear{Bachtis, Aarts, and Lucini}{Bachtis et~al.}{2020}]{bachtis2020mapping}
Bachtis, D., G.~Aarts, and B.~Lucini. 2020.
\newblock Mapping distinct phase transitions to a neural network.
\newblock {\em Physical Review E\/}~{\em 102\/}(5): 053306 .

\bibitem[\protect\citeauthoryear{Banerjee}{Banerjee}{1992}]{banerjee1992simple}
Banerjee, A.V. 1992.
\newblock A simple model of herd behavior.
\newblock {\em The quarterly journal of economics\/}~{\em 107\/}(3): 797--817 .

\bibitem[\protect\citeauthoryear{Barrat, Barthelemy, and Vespignani}{Barrat et~al.}{2008}]{barrat2008dynamical}
Barrat, A., M.~Barthelemy, and A.~Vespignani. 2008.
\newblock {\em Dynamical processes on complex networks}.
\newblock Cambridge: Cambridge University Press.

\bibitem[\protect\citeauthoryear{Bikhchandani, Hirshleifer, and Welch}{Bikhchandani et~al.}{1992}]{bikhchandani1992theory}
Bikhchandani, S., D.~Hirshleifer, and I.~Welch. 1992.
\newblock A theory of fads, fashion, custom, and cultural change as informational cascades.
\newblock {\em Journal of political Economy\/}~{\em 100\/}(5): 992--1026 .

\bibitem[\protect\citeauthoryear{Borge-Holthoefer, Banos, Gonz{\'a}lez-Bail{\'o}n, and Moreno}{Borge-Holthoefer et~al.}{2013}]{borge2013cascading}
Borge-Holthoefer, J., R.A. Banos, S.~Gonz{\'a}lez-Bail{\'o}n, and Y.~Moreno. 2013.
\newblock Cascading behaviour in complex socio-technical networks.
\newblock {\em Journal of Complex Networks\/}~{\em 1\/}(1): 3--24 .

\bibitem[\protect\citeauthoryear{Bouallegue}{Bouallegue}{2017}]{bouallegue2017analyse}
Bouallegue, O. 2017.
\newblock {\em Analyse {\'e}conomique des r{\'e}volutions: Cas de la r{\'e}volution Tunisienne}.
\newblock Ph.\ D. thesis, Universit{\'e} Montpellier.

\bibitem[\protect\citeauthoryear{Castellano, Marsili, and Vespignani}{Castellano et~al.}{2000}]{castellano2000nonequilibrium}
Castellano, C., M.~Marsili, and A.~Vespignani. 2000.
\newblock Nonequilibrium phase transition in a model for social influence.
\newblock {\em Physical Review Letters\/}~{\em 85\/}(16): 3536 .

\bibitem[\protect\citeauthoryear{Chenavaz}{Chenavaz}{2012}]{chenavaz2012printemps}
Chenavaz, R. 2012.
\newblock Printemps arabe: les r{\'e}seaux sociaux suffisent-ils {\`a} renverser un r{\'e}gime?
\newblock {\em L’Obs\/}~3 .

\bibitem[\protect\citeauthoryear{Cohen}{Cohen}{2011}]{cohen2011facebook}
Cohen, R. 2011.
\newblock Facebook and arab dignity.
\newblock {\em The New York Times\/}~24 .

\bibitem[\protect\citeauthoryear{Dabashi}{Dabashi}{2012}]{dabashi2012arab}
Dabashi, H. 2012.
\newblock {\em The Arab spring: The end of postcolonialism}.
\newblock London: Bloomsbury Publishing.

\bibitem[\protect\citeauthoryear{de~Arruda, Petri, and Moreno}{de~Arruda et~al.}{2020}]{de2020social}
de~Arruda, G.F., G.~Petri, and Y.~Moreno. 2020.
\newblock Social contagion models on hypergraphs.
\newblock {\em Physical Review Research\/}~{\em 2\/}(2): 023032 .

\bibitem[\protect\citeauthoryear{Deffuant, Neau, Amblard, and Weisbuch}{Deffuant et~al.}{2000}]{deffuant2000mixing}
Deffuant, G., D.~Neau, F.~Amblard, and G.~Weisbuch. 2000.
\newblock Mixing beliefs among interacting agents.
\newblock {\em Advances in Complex Systems\/}~{\em 3\/}(01n04): 87--98 .

\bibitem[\protect\citeauthoryear{Dosi}{Dosi}{2023}]{dosi2023foundations}
Dosi, G. 2023.
\newblock {\em The foundations of complex evolving economies: Part one: Innovation, organization, and industrial dynamics}.
\newblock Oxford: Oxford University Press.

\bibitem[\protect\citeauthoryear{Dunbar}{Dunbar}{1992}]{dunbar1992neocortex}
Dunbar, R.I. 1992.
\newblock Neocortex size as a constraint on group size in primates.
\newblock {\em Journal of human evolution\/}~{\em 22\/}(6): 469--493 .

\bibitem[\protect\citeauthoryear{Filiu}{Filiu}{2011}]{filiu2011revolution}
Filiu, J.P. 2011.
\newblock La r{\'e}volution arabe.
\newblock {\em Dix lecons sur le soulevement democratique. P\/}: 34--35 .

\bibitem[\protect\citeauthoryear{Gelvin}{Gelvin}{2015}]{gelvin2015arab}
Gelvin, J.L. 2015.
\newblock {\em The Arab uprisings: What everyone needs to know}.
\newblock Oxford: Oxford University Press.

\bibitem[\protect\citeauthoryear{Goldenfeld}{Goldenfeld}{2018}]{goldenfeld1992lectures}
Goldenfeld, N.D. 2018.
\newblock {\em Lectures On Phase Transitions And The Renormalization Group}.
\newblock Boca Raton: CRC Press.

\bibitem[\protect\citeauthoryear{H{\'e}bert-Dufresne and Althouse}{H{\'e}bert-Dufresne and Althouse}{2015}]{hebert2015complex}
H{\'e}bert-Dufresne, L. and B.M. Althouse. 2015.
\newblock Complex dynamics of synergistic coinfections on realistically clustered networks.
\newblock {\em Proceedings of the National Academy of Sciences\/}~{\em 112\/}(33): 10551--10556 .

\bibitem[\protect\citeauthoryear{Hill}{Hill}{1955}]{hill1993revolution}
Hill, C. 1955.
\newblock {\em The English Revolution, 1640}.
\newblock London: Lawrence $\&$ Wishart.

\bibitem[\protect\citeauthoryear{Holme and Newman}{Holme and Newman}{2006}]{holme2006nonequilibrium}
Holme, P. and M.E.J. Newman. 2006.
\newblock Nonequilibrium phase transition in the coevolution of networks and opinions.
\newblock {\em Physical Review E\/}~{\em 74\/}(5): 056108 .

\bibitem[\protect\citeauthoryear{Howard, Duffy, Freelon, Hussain, Mari, and Maziad}{Howard et~al.}{2011}]{howard2011opening}
Howard, P., A.~Duffy, D.~Freelon, M.M. Hussain, W.~Mari, and M.~Maziad. 2011.
\newblock Opening closed regimes: what was the role of social media during the arab spring? .

\bibitem[\protect\citeauthoryear{Iniguez, Kert{\'e}sz, Kaski, and Barrio}{Iniguez et~al.}{2009}]{iniguez2009opinion}
Iniguez, G., J.~Kert{\'e}sz, K.K. Kaski, and R.A. Barrio. 2009.
\newblock Opinion and community formation in coevolving networks.
\newblock {\em Physical Review E—Statistical, Nonlinear, and Soft Matter Physics\/}~{\em 80\/}(6): 066119 .

\bibitem[\protect\citeauthoryear{Janne}{Janne}{1960}]{janne1960modele}
Janne, H. 1960.
\newblock Un mod{\`e}le th{\'e}orique du ph{\'e}nom{\`e}ne r{\'e}volutionnaire?
\newblock In {\em Annales. Histoire, Sciences Sociales}, Volume~15, pp.\  1138--1154. Cambridge University Press.

\bibitem[\protect\citeauthoryear{Joshi}{Joshi}{2011}]{joshi2011arab}
Joshi, S. 2011.
\newblock Arab spring: Nature of armies decisive in revolutions, british broadcasting corporation.
\newblock {\em Middle East\/}~28 .

\bibitem[\protect\citeauthoryear{Kinouchi and Copelli}{Kinouchi and Copelli}{2006}]{kinouchi2006optimal}
Kinouchi, O. and M.~Copelli. 2006.
\newblock Optimal dynamical range of excitable networks at criticality.
\newblock {\em Nature physics\/}~{\em 2\/}(5): 348--351 .

\bibitem[\protect\citeauthoryear{Kuramoto}{Kuramoto}{1975}]{kuramosto1975self}
Kuramoto, Y. 1975.
\newblock Self-entrainment of a population of coupled non-linear oscillators.
\newblock In H.~Araki (Ed.), {\em International Symposium on Mathematical Problems in Theoretical Physics, Lecture Notes in Physics No. 30}, New York, pp.\  420. Springer.

\bibitem[\protect\citeauthoryear{Kuran}{Kuran}{1989}]{kuran1989sparks}
Kuran, T. 1989.
\newblock Sparks and prairie fires: A theory of unanticipated political revolution.
\newblock {\em Public choice\/}~{\em 61\/}(1): 41--74 .

\bibitem[\protect\citeauthoryear{Labrousse}{Labrousse}{1949}]{labrousse1949comment}
Labrousse, E. 1949.
\newblock {\em Comment naissent les r{\'e}volutions: 1848, 1830, 1789}.

\bibitem[\protect\citeauthoryear{Le~Bon}{Le~Bon}{1995}]{le1995psikhologiya}
Le~Bon, G. 1995.
\newblock Psikhologiya narodov i mass [psychologie des foules].
\newblock {\em St. Petersburg, Maket Publ\/} .

\bibitem[\protect\citeauthoryear{Lecomte}{Lecomte}{2011}]{lecomte2011tunisian}
Lecomte, R. 2011.
\newblock Tunisian revolution and the internet: The role of social media.
\newblock {\em R{\'e}volution tunisienne et Internet: le r{\^o}le des m{\'e}dias sociaux. L'Ann{\'e}e du Maghreb\/}~7: 389--418 .

\bibitem[\protect\citeauthoryear{Levy}{Levy}{2005}]{levy2005social}
Levy, M. 2005.
\newblock Social phase transitions.
\newblock {\em Journal of Economic Behavior \& Organization\/}~{\em 57\/}(1): 71--87 .

\bibitem[\protect\citeauthoryear{Lucas~Jr}{Lucas~Jr}{1972}]{lucas1972expectations}
Lucas~Jr, R.E. 1972.
\newblock Expectations and the neutrality of money.
\newblock {\em Journal of economic theory\/}~{\em 4\/}(2): 103--124 .

\bibitem[\protect\citeauthoryear{Malkov, Grinin, Grinin, Musieva, and Korotayev}{Malkov et~al.}{2023}]{malkov2023modeling}
Malkov, S., L.~Grinin, A.~Grinin, J.~Musieva, and A.~Korotayev. 2023.
\newblock Modeling social self-organization and historical dynamics: Global phase transitions, {\em Reconsidering the Limits to Growth: A Report to the Russian Association of the Club of Rome},  387--417. Springer.

\bibitem[\protect\citeauthoryear{Marro and Dickman}{Marro and Dickman}{1999}]{marro1999noequilibrium}
Marro, J. and R.~Dickman. 1999.
\newblock {\em Nonequilibrium Phase Transitions in Lattice Models}.
\newblock Cambridge: Cambridge University Press.

\bibitem[\protect\citeauthoryear{Marsili, Vega-Redondo, and Slanina}{Marsili et~al.}{2004}]{marsili2004rise}
Marsili, M., F.~Vega-Redondo, and F.~Slanina. 2004.
\newblock The rise and fall of a networked society: A formal model.
\newblock {\em Proceedings of the National Academy of Sciences\/}~{\em 101\/}(6): 1439--1442 .

\bibitem[\protect\citeauthoryear{Martinez}{Martinez}{2020}]{martinez2020state}
Martinez, L. 2020.
\newblock {\em The State in North Africa: After the Arab Uprisings}.
\newblock New York: Oxford University Press, USA.

\bibitem[\protect\citeauthoryear{Marwell and Oliver}{Marwell and Oliver}{1993}]{marwell1993critical}
Marwell, G. and P.~Oliver. 1993.
\newblock {\em The critical mass in collective action}.
\newblock Cambridge, UK: Cambridge University Press.

\bibitem[\protect\citeauthoryear{McDonald and Moffitt}{McDonald and Moffitt}{1980}]{mcdonald1980uses}
McDonald, J.F. and R.A. Moffitt. 1980.
\newblock The uses of tobit analysis.
\newblock {\em The review of economics and statistics\/}: 318--321 .

\bibitem[\protect\citeauthoryear{Murase, Jo, T{\"o}r{\"o}k, Kert{\'e}sz, and Kaski}{Murase et~al.}{2015}]{murase2015modeling}
Murase, Y., H.H. Jo, J.~T{\"o}r{\"o}k, J.~Kert{\'e}sz, and K.~Kaski. 2015.
\newblock Modeling the role of relationship fading and breakup in social network formation.
\newblock {\em PloS ONE\/}~{\em 10\/}(7): e0133005 .

\bibitem[\protect\citeauthoryear{Murase, Jo, T{\"o}r{\"o}k, Kert{\'e}sz, and Kaski}{Murase et~al.}{2019}]{murase2019structural}
Murase, Y., H.H. Jo, J.~T{\"o}r{\"o}k, J.~Kert{\'e}sz, and K.~Kaski. 2019.
\newblock Structural transition in social networks: The role of homophily.
\newblock {\em Scientific Reports\/}~{\em 9\/}(1): 4310 .

\bibitem[\protect\citeauthoryear{Nelson and Winter}{Nelson and Winter}{1985}]{nelson1985evolutionary}
Nelson, R.R. and S.G. Winter. 1985.
\newblock {\em An evolutionary theory of economic change}.
\newblock Cambridge, USA: Harvard University Press.

\bibitem[\protect\citeauthoryear{Newman}{Newman}{2003}]{newman2003structure}
Newman, M.E. 2003.
\newblock The structure and function of complex networks.
\newblock {\em SIAM review\/}~{\em 45\/}(2): 167--256 .

\bibitem[\protect\citeauthoryear{Olsson-Yaouzis}{Olsson-Yaouzis}{2012}]{olsson2012evolutionary}
Olsson-Yaouzis, N. 2012.
\newblock An evolutionary dynamic of revolutions.
\newblock {\em Public Choice\/}~151: 497--515 .

\bibitem[\protect\citeauthoryear{Oualdi, Pag{\`e}s-El~Karoui, and Verdeil}{Oualdi et~al.}{2014}]{oualdi2014ondes}
Oualdi, M., D.~Pag{\`e}s-El~Karoui, and C.~Verdeil. 2014.
\newblock {\em Les ondes de choc des r{\'e}volutions arabes}, Volume~36.
\newblock Paris: Presses de l’Ifpo.

\bibitem[\protect\citeauthoryear{Palfrey and Rosenthal}{Palfrey and Rosenthal}{1984}]{palfrey1984participation}
Palfrey, T.R. and H.~Rosenthal. 1984.
\newblock Participation and the provision of discrete public goods: a strategic analysis.
\newblock {\em Journal of public Economics\/}~{\em 24\/}(2): 171--193 .

\bibitem[\protect\citeauthoryear{Pastor-Satorras, Castellano, Van~Mieghem, and Vespignani}{Pastor-Satorras et~al.}{2015}]{pastor2015epidemic}
Pastor-Satorras, R., C.~Castellano, P.~Van~Mieghem, and A.~Vespignani. 2015.
\newblock Epidemic processes in complex networks.
\newblock {\em Reviews of Modern Physics\/}~{\em 87\/}(3): 925--979 .

\bibitem[\protect\citeauthoryear{Perc}{Perc}{2016}]{perc2016phase}
Perc, M. 2016.
\newblock Phase transitions in models of human cooperation.
\newblock {\em Physics Letters A\/}~{\em 380\/}(36): 2803--2808 .

\bibitem[\protect\citeauthoryear{{Petit Larousse}}{{Petit Larousse}}{1986}]{larousse1986dictionnaire}
{Petit Larousse}. 1986.
\newblock Dictionnaire encyclop{\'e}dique illustr{\'e}.

\bibitem[\protect\citeauthoryear{Roemer}{Roemer}{1985}]{roemer1985rationalizing}
Roemer, J.E. 1985.
\newblock Rationalizing revolutionary ideology.
\newblock {\em Econometrica: Journal of the Econometric Society\/}: 85--108 .

\bibitem[\protect\citeauthoryear{Scarpetta, Apicella, Minati, and De~Candia}{Scarpetta et~al.}{2018}]{scarpetta2018hysteresis}
Scarpetta, S., I.~Apicella, L.~Minati, and A.~De~Candia. 2018.
\newblock Hysteresis, neural avalanches, and critical behavior near a first-order transition of a spiking neural network.
\newblock {\em Physical Review E\/}~{\em 97\/}(6): 062305 .

\bibitem[\protect\citeauthoryear{Skardal and Arenas}{Skardal and Arenas}{2020}]{skardal2020higher}
Skardal, P.S. and A.~Arenas. 2020.
\newblock Higher order interactions in complex networks of phase oscillators promote abrupt synchronization switching.
\newblock {\em Communications Physics\/}~{\em 3\/}(1): 218 .

\bibitem[\protect\citeauthoryear{Stone}{Stone}{2017}]{stone2017theories}
Stone, L. 2017.
\newblock Theories of revolution, {\em Revolutionary Guerrilla Warfare},  27--46. London: Routledge.

\bibitem[\protect\citeauthoryear{Strogatz and Mirollo}{Strogatz and Mirollo}{1991}]{strogatz1991stability}
Strogatz, S.H. and R.E. Mirollo. 1991.
\newblock Stability of incoherence in a population of coupled oscillators.
\newblock {\em Journal of Statistical Physics\/}~63: 613--635 .

\bibitem[\protect\citeauthoryear{Tullock}{Tullock}{1974}]{tullock1974social}
Tullock, G. 1974.
\newblock The social dilemma: The economics of war and revolution.
\newblock {\em (No Title)\/} .

\bibitem[\protect\citeauthoryear{Tullock and Rowley}{Tullock and Rowley}{2005}]{tullock2005social}
Tullock, G. and C.K. Rowley. 2005.
\newblock The social dilemma: Of autocracy, revolution, coup d'etat, and war.
\newblock {\em (No Title)\/} .

\bibitem[\protect\citeauthoryear{Wang, Siebenh{\"u}hner, Arnulfo, Myrov, Nobili, Breakspear, Palva, and Palva}{Wang et~al.}{2023}]{wang2023critical}
Wang, S.H., F.~Siebenh{\"u}hner, G.~Arnulfo, V.~Myrov, L.~Nobili, M.~Breakspear, S.~Palva, and J.M. Palva. 2023.
\newblock Critical-like brain dynamics in a continuum from second-to first-order phase transition.
\newblock {\em Journal of Neuroscience\/}~{\em 43\/}(45): 7642--7656 .

\bibitem[\protect\citeauthoryear{Yeomans}{Yeomans}{1992}]{yeomans1992statistical}
Yeomans, J.M. 1992.
\newblock {\em Statistical mechanics of phase transitions}.
\newblock Oxford: Clarendon Press.

\end{thebibliography}
\end{document}